\documentclass[twocolumn,prl,superscriptaddress]{revtex4}
\usepackage{graphicx}
\usepackage[hypertex]{hyperref}
\usepackage{times}

\begin{document}

\title{Viable Supersymmetry and Leptogenesis with Anomaly Mediation}

\author{Masahiro Ibe}
\affiliation{Department of Physics, University of Tokyo, Tokyo
  113-0033, Japan} 
\author{Ryuichiro Kitano}
\affiliation{School of Natural Sciences, Institute for Advanced
Study, Princeton, NJ 08540}
\author{Hitoshi Murayama}
\thanks{On leave of absence from Department of Physics,
University of California, Berkeley, CA 94720.}
\affiliation{School of Natural Sciences, Institute for Advanced
Study, Princeton, NJ 08540}
\author{Tsutomu Yanagida}
\affiliation{Department of Physics, University of Tokyo, Tokyo
  113-0033, Japan} 

\date{March 17, 2004}


\begin{abstract}
  The seesaw mechanism that explains the small neutrino masses comes
  naturally with supersymmetric (SUSY) grand unification and
  leptogenesis.  However, the framework suffers from the SUSY flavor
  and CP problems, and has a severe cosmological gravitino problem.
  We propose anomaly mediation as a simple solution to all these
  problems, which is viable once supplemented by the $D$-terms for
  $U(1)_Y$ and $U(1)_{B-L}$.  Even though the right-handed neutrino
  mass explicitly breaks $U(1)_{B-L}$ and hence reintroduces the
  flavor problem, we show that it lacks the logarithmic enhancement
  and poses no threat to the framework.  The thermal leptogenesis is
  then made easily consistent with the gravitino constraint.
\end{abstract}
\maketitle


The past several years have seen revolutionary progress in neutrino
physics.  The atmospheric neutrinos showed the first convincing
evidence for neutrino oscillation in the SuperKamiokande experiment
\cite{Fukuda:1998mi}.  The long-standing solar neutrino problem since
1960s was shown to be due to the neutrino flavor conversion by the SNO
experiment \cite{Ahmad:2002jz}, and all possibilities other than
neutrino oscillation have been excluded by the KamLAND experiment
\cite{Eguchi:2002dm}.  All these experiments suggest finite but
extremely small neutrino mass in the sub-electronvolt range, more than
a million times smaller than the smallest particle mass known before,
namely the electron mass.  

The standard framework to understand the newly discovered neutrino
masses and mixings is the seesaw mechanism \cite{seesaw}, which comes
naturally with the supersymmetric (SUSY) grand-unified theories (GUT).
Here, SUSY plays a dual role: it stabilizes the hierarchy and makes
the gauge coupling constants unify.  Furthermore, the seesaw mechanism
predicts new heavy particles, right-handed neutrinos, whose decay can
potentially produce the baryon asymmetry of the universe
\cite{Fukugita:1986hr}.  This possibility is called leptogenesis,
which requires $T_{RH} > 4 \times 10^9$~GeV to generate the observed
baryon asymmetry \cite{Giudice:2003jh} \footnote{A high degree of
  degeneracy among right-handed neutrinos may resonantly enhance the
  baryon asymmetry and may allow for lighter right-handed neutrinos
  \cite{Pilaftsis:2003gt}.}.  The combination of seesaw, SUSY GUT, and
leptogenesis is further supplemented by the possibility of SUSY dark
matter, which predicts the cosmic abundance of the dark matter
particle in the right ballpark.

There are, however, severe problems with this attractive framework.
SUSY tends to induce unacceptably large flavor-changing and
CP-violating effects; SUSY flavor and CP problems \cite{gabbiani}.
Generic SUSY parameters imply a lower limit on the masses in excess of
100~TeV, making supersymmetry not a viable mechanism to stabilize the
electroweak scale $m_Z = 91$~GeV.  It is customary to make an {\it ad
  hoc}\/ assumption of the universal scalar mass at the GUT- or
Planck-scale to avoid the flavor problem.  However, the rates of
Lepton Flavor Violation (LFV) processes are typically predicted to be
too high in SUSY-GUT models based on the seesaw mechanism and flavor
symmetries even with the universal scalar mass \cite{Sato:2000zh}.  At
the same time, SUSY predicts the existence of the gravitino, the
superpartner of the graviton.  Once it is produced in early universe,
it decays only by the gravitational interaction and hence slowly,
upsetting the success of BBN \cite{Pagels:ke}.  The gravitino yield is
larger for higher reheating temperatures.  A detailed analysis
including the hadronic decay of gravitino shows a very tight upper
limit $T_{RH} \ll 10^6$~GeV \cite{Kawasaki:2004yh} for $m_{3/2} =
0.1$--1~TeV as commonly assumed in the literature.  Therefore the
leptogenesis appears incompatible with the gravitino problem.

There had been suggestions to achieve leptogenesis at a relatively low
temperature (for a compilation of proposals, see
\cite{Davidson:2002qv}).  In particular, coherent oscillation of
right-handed scalar neutrino \cite{Murayama:1993em} had been
considered a natural possibility, and it may even be the inflaton
\cite{Murayama:1992ua}.  However, this proposal requires the reheating
temperature to be above $10^6$~GeV \cite{Hamaguchi:2001gw} which is
still in conflict with the gravitino problem.  More recent suggestions
include the gravitino LSP \cite{Fujii:2003nr}, but even this case is
getting tightly constrained.

In this Letter, we revisit these problems and propose a simple
solution: anomaly-mediated supersymmetry breaking
\cite{Randall:1998uk,Giudice:1998xp}.  It predicts a heavy gravitino
that decays before BBN and makes leptogenesis viable.  It solves the
flavor and CP problems automatically if supplemented by the
UV-insensitive $D$-terms \cite{Arkani-Hamed:2000xj}.  Even though the
strict version of the UV-insensitive anomaly mediation requires $B-L$
conservation, it was pointed out that the seesaw mechanism can be used
with minimal flavor-changing effects \cite{Arkani-Hamed:2000xj}.
However, the details of reintroduced flavor-changing effects had not
been worked out.  We present consequences of the seesaw mechanism on
LFV.  We find that the LFV effects lack logarithmic enhancements
unlike in conventional supergravity-based scenarios and hence are
easily compatible with current limits.  Therefore this framework
preserves all virtues of anomaly mediation to solve the SUSY flavor
and CP problems and makes the thermal leptogenesis viable.  On the
other hand, the small LFV may lead to a signature observable in the
near future.

Let us set up notations to discuss seesaw mechanism with SUSY GUT.
The relevant part of the superpotential is
\begin{equation}
  \label{eq:W}
  W = h_{i\alpha} L_i N_\alpha H_u + \frac{1}{2} M_\alpha N_\alpha N_\alpha.
\end{equation}
We use the basis where the right-handed neutrino masses $M_\alpha$ are
diagonal and real positive.  The light neutrino mass is then obtained
by integrating out $N_\alpha$, $(m_\nu)_{ij} = \sum_\alpha h_{i\alpha}
h_{j\alpha} \langle H_u\rangle^2 / M_\alpha$.  Therefore the light
neutrino masses are suppressed relative to the other quark and lepton
masses by the inverse power of $M_\alpha$.  In order to obtain the
heaviest mass $m_3 \gtrsim \sqrt{\Delta m^2_{\rm atm}} \simeq
0.05$~eV, we find $M_3 \lesssim 6 \times 10^{14}$~GeV, which can be
induced from the grand-unification scale $M_{GUT} \simeq 2 \times
10^{16}$~GeV.

How do we solve the SUSY flavor and CP problems, while make the
leptogenesis consistent with the gravitino problem?  There is a
promising mechanism to make the whole framework consistent.  Anomaly
mediation of SUSY breaking \cite{Randall:1998uk,Giudice:1998xp} induces
the SUSY breaking effects from the superconformal anomaly, and hence
they are determined solely by physics at the energy scale of interest.
When applied to the SUSY standard model, it automatically solves the
serious flavor and CP problems.  Because the anomaly is a quantum effect
and hence loop-suppressed, the typical SUSY masses are smaller than the
gravitino mass by $(4\pi)^2$, implying that the gravitino is heavy,
$m_{3/2} \simeq 100$~TeV.  Such a large mass allows the gravitino to
decay before BBN, and hence the gravitino is harmless.  The only
constraint is that stable Lightest Supersymmetric Particle (LSP) from
the gravitino decay does not provide too much dard matter of the
universe. It requires \cite{Kawasaki:1994af}
\begin{equation}
  \label{eq:TRH}
  T_{RH} \leq 2.1\times 10^{10}~{\rm GeV} \left( \frac{m_{LSP}}{\rm
  100~GeV}\right)^{-1}.
\end{equation}
This constraint can be satisfied together with the requirement for the
thermal leptogenesis $T_{RH} > 4 \times 10^9$~GeV.
The upper bound on the LSP mass, $m_{\rm LSP} \lesssim 500$~GeV, is
derived from the consistency between those two requirements, by assuming
there is no significant annihilation after the gravitino decays.

Despite these attractive features, anomaly mediation has not been used
widely in the literature because of several initial problems.  The
slepton mass-squared comes out negative, breaking the electromagnetism
spontaneously.  Many fixes proposed in the literature
\cite{Pomarol:1999ie,Chacko:1999am,Katz:1999uw,Bagger:1999rd,Kaplan:2000jz,Chacko:2001jt,Okada:2002mv,Nelson:2002sa,Anoka:2003kn}
unfortunately spoils the UV insensitivity and hence reintroduces the
flavor and CP problems, unless $R$-parity is violated
\cite{Allanach:2000gu}.  On the other hand, the addition of $D$-terms
for $U(1)_Y$ and $U(1)_{B-L}$ can make the slepton mass-squared
positive \cite{Jack:2000cd}, and furthermore the UV insensitivity is
preserved \cite{Arkani-Hamed:2000xj}.  The viable electroweak symmetry
breaking was demonstrated only recently \cite{Kitano:2004zd} which
goes extremely well with the low-energy limit of the Minimal Fat Higgs
Model \cite{Harnik:2003rs}.  Even though the original setting relied
on extra dimensions \cite{Randall:1998uk}, it can now be constructed
in a purely four-dimensional setting \cite{Luty:2001jh} together with
the required $D$-terms \cite{Harnik:2002et}.  Therefore, anomaly
mediation can be finally regarded as a consistent and viable framework
of supersymmetry breaking.

The $U(1)_{B-L}$ gauge invariance is broken at some high scale, and its
only remnant is its $D$-term $V_{B-L} = \theta^2 \bar{\theta}^2 D_{B-L}$
and an accidental global non-anomalous $U(1)_{B-L}$ symmetry to ensure
the UV insensitivity.  The K\"ahler potential $\int d^4 \theta \phi_i^*
e^{q_i V_{B-L}} \phi_i$ for a matter field of $B-L$ charge $q_i$ gives a
contribution to its mass-squared $(m_i^2)_D = - q_i D_{B-L}$ (there is also
Fayet--Illiopoulos term for $U(1)_Y$ that is not relevant in this
paper).  On the other hand, the seesaw mechanism breaks $B-L$ explicitly
by the Majorana mass of right-handed neutrinos, and reintroduces the
flavor violation in a highly controlled fashion.  This effect had not
been worked out quantitatively in the original work
\cite{Arkani-Hamed:2000xj}, and we study it in detail in this Letter.

We first derive the expression of the threshold corrections from the
right-handed neutrinos.
The correction to the slepton mass matrices due to the $B-L$ breaking
in right-handed Majorana masses can be worked out using the ``analytic
continuation into superspace'' \cite{Arkani-Hamed:1998kj} used
extensively in anomaly mediation
\cite{Pomarol:1999ie,Arkani-Hamed:2000xj,Boyda:2001nh}.  $M_\alpha$ in
Eq.~(\ref{eq:W}) are the only $B-L$ violating parameters in the
theory.  Using the fictitious gauge invariance $N_\alpha \rightarrow
e^{i q_\alpha \Lambda} N_\alpha$, $e^{V_{B-L}} \rightarrow e^{i
  \Lambda^*} e^{V_{B-L}} e^{-i\Lambda}$, and $M_\alpha \rightarrow
M_\alpha e^{-i 2 q_{\alpha} \Lambda}$, it is clear that the only
combination that can appear in the low-energy theory is $M_\alpha^*
e^{-2 q_\alpha V_{B-L}} M_\alpha$.  It appears in the K\"ahler
potential for the left-handed leptons $L_i$ and $H_u$ at the one-loop
level
\begin{equation}
  \label{eq:Zij}
  Z_{ij} = Z_{ij}^0 - \sum_\alpha \frac{h_{i\alpha}h_{j\alpha}^*}{(4\pi)^2}
  \log \frac{M_\alpha^* e^{-2q_\alpha V_{B-L}} M_\alpha}{\mu^2},
\end{equation}
$q_\alpha=+1$ is the $B-L$ charge of the $N_\alpha$ superfield.  $\mu$
is the renormalization scale.  The wave-function renormalization
factor $Z_{ij}^0$ is for the case of massless right-handed neutrinos
and exact $U(1)_{B-L}$ invariance, and hence gives a fully UV
insensitive supersymmetry breaking.  The second term subtracts the
contribution of the right-handed neutrinos between their mass
thresholds $M_\alpha$ and the renormalization scale $\mu \ll
M_\alpha$.  Here, the vector superfield $V_{B-L}$ contains the
required $D$-term, and makes $M_\alpha$ dependence invariant under the
(spurious) $U(1)_{B-L}$ symmetry.  It induces the correction to the
left-handed slepton mass-squared matrix,
\begin{equation}
  \label{eq:m2ij}
  \Delta m^2_{ij} = - 2 \sum_\alpha 
  \frac{h_{i\alpha}h_{j\alpha}^*}{(4\pi)^2} D_{B-L}.
\end{equation}
Note that there is no flavor-violating correction to the right-handed
sleptons.  Corresponding formula in the minimal supergravity (mSUGRA)
that assumes the universal scalar mass is \cite{Hisano:1995nq}
\begin{equation}
  \label{eq:mSUGRA}
  \Delta m^2_{ij} = - \sum_\alpha \frac{h_{i\alpha}h_{j\alpha}^*}{(4\pi)^2}
  (3m_0^2 + A_0^2) \log \frac{\Lambda_{UV}^2}{M_\alpha^2}, 
\end{equation}
where $m_0$ is the universal scalar mass at the ultraviolet cutoff
$\Lambda_{UV}$ and $A_0$ the universal trilinear coupling.

There are several remarkable aspects in Eq.~(\ref{eq:m2ij}).  First,
there is no logarithmic enhancement.  In contrast, the usual
supergravity theories are UV sensitive and hence the contributions of
the right-handed neutrinos are enhanced by $\log
(\Lambda_{UV}/M_\alpha)$ as seen in Eq.~(\ref{eq:mSUGRA}).  Therefore
the size of possible LFV is under a much stronger control in the
anomaly mediation.
Related to the aspect, Eq.~(\ref{eq:m2ij}) is independent of physics
above the right-handed mass scale since the contribution originates from
the threshold effect of $N_\alpha$ as we see below in the explicit
diagrammatic calculation. In the mSUGRA, in contrast, we need to know
the theory above $M_\alpha$, especially above $M_{GUT}$, to calculate
$\Delta m^2_{ij}$ by carrying out the integration of the renormalization
group equations with initial conditions given at the Planck scale.
Second, the corrections do not depend on the mass of the right-handed
neutrinos $M_\alpha$ explicitly (but implicitly through $h_{i\alpha}$
once the light neutrino masses are held fixed).
Once we observe LFV processes and measure the branching ratios, the
simple structure of $\Delta m^2_{ij}$ in Eq.~(\ref{eq:m2ij}) enables
us to extract easily the the Yukawa matrix $h_{i \alpha}$ which has
important information on the origin of the large mixing among
neutrinos, {\it i.e.}, whether the large mixing comes from the
left-handed lepton sector, the right-handed neutrino sector, or both
sectors.
%
Third, because the corrections are suppressed by the one-loop factor
relative to the leading anomaly-mediated contributions thanks to the
absence of the logarithmic enhancement, the leading-order trajectory
of the soft SUSY breaking parameters is still strictly that of the UV
insensitive anomaly mediation.  Finally, even if there is a
flavor-violating interaction between quarks and leptons below
$M_\alpha$ ({\it i.e.}\/, leptoquarks), the induced flavor-violation
in quarks is at most of the order of $\Delta m^2_{ij}$ above, while
the squarks are heavier than the typical size of $D_{B-L}$ by a factor
of $g_s^4/g^4$.  Therefore it will not pose a serious threat.

There is a corresponding correction to the Higgs mass
\begin{equation}
  \label{eq:m2Hu}
  \Delta m^2_{H_u} = - 2 \sum_{i,\alpha}
  \frac{h_{i\alpha}h_{i\alpha}^*}{(4\pi)^2} D_{B-L}.
\end{equation}
However, this is a loop-suppressed effect relative to the leading
anomaly-mediated contribution, and does not lead to flavor-violating
effects.  It can be safely ignored for all practical purposes.

\begin{figure}[t]
\includegraphics[width=7.5cm]{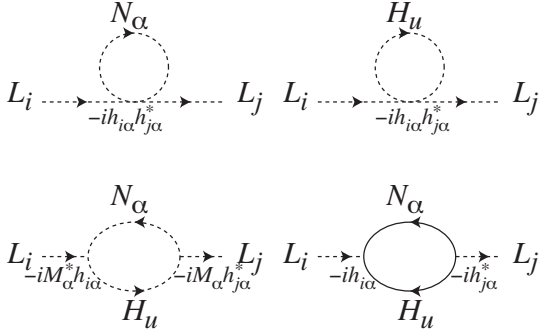} 
\caption{
The Feynman diagrams which contribute to the slepton soft masses.
}
\label{fig:feynman}
\end{figure}

Because the absence of the logarithmic enhancement is a striking
result, it is useful to examine it with the conventional Feynman
diagrams shown in Fig.~\ref{fig:feynman}.  We use regularization by
dimensional reduction (DRED) in $D=4-2\epsilon$ dimensions to perform
loop integrals.  Because the anomaly-mediated pieces due to the
neutrino Yukawa couplings are canceled exactly by the threshold
corrections \cite{Arkani-Hamed:1998kj,Boyda:2001nh}, we only consider
$D$-term contributions to the scalar masses-squared, $m_\alpha^2$ for
$N_\alpha$, $m_{H_u}^2$ for $H_u$, and $m^2_{i}$ for $L_i$ (of course
$m_{H_u}^2 \propto q_{H_u} =0$, but we retain it for the clarity of
presentation).  The sum of all boson loops at $q^2 = 0$ gives
\begin{eqnarray}
  \label{eq:bosonic}
  \lefteqn{
    \frac{ih_{i\alpha} h^*_{j\alpha}}{(4\pi)^{2-\epsilon}} 
    \Biggl\{ - 2m_\alpha^2 } \nonumber \\
  & & 
    +(2M_\alpha^2 + m_{H_u}^2 + m_\alpha^2)
    \left( \frac{1}{\epsilon} + 1 -\gamma -
      \log\frac{M_\alpha^2}{\mu^2} \right) \Biggr\},
\end{eqnarray}
while the fermion loop gives the correction at $q^2 = 0$,
\begin{equation}
  \label{eq:fermionic}
  \frac{i h_{i\alpha} h^*_{j\alpha}}{(4\pi)^{2-\epsilon}}
  (-2M_\alpha^2)
  \left( \frac{1}{\epsilon} + 1 -\gamma -
    \log\frac{M_\alpha^2}{\mu^2} \right).
\end{equation}
Finally the leading $q^2$ dependence of the two-point function is
\begin{equation}
  \label{eq:wavefunction}
   q^2 \frac{i h_{i\alpha} h^*_{j\alpha}}{(4\pi)^{2-\epsilon}}
  \left( \frac{1}{\epsilon} + 1 -\gamma -
      \log\frac{M_\alpha^2}{\mu^2} \right),
\end{equation}
which gives the wave function renormalization factor.
Putting them together, the final correction to the slepton masses-squared is 
\begin{eqnarray}
  \label{eq:final}
  \lefteqn{
    -i \Delta m^2_{ij} 
    = \frac{i h_{i\alpha} h^*_{j\alpha}}{(4\pi)^{2-\epsilon}}
    \Biggl\{ - 2 m_\alpha^2 
    } \nonumber \\
    & &
    + (m_{H_u}^2 + m_\alpha^2 + m_i^2)
      \left( \frac{1}{\epsilon} + 1 -\gamma -
        \log\frac{M_\alpha^2}{\mu^2} \right)\Biggr\}.
\end{eqnarray}
Because the $D$-term contributions satisfy $m_{H_u}^2 + m_\alpha^2 +
m_i^2 = -(q_{H_u} + q_\alpha + q_i) D_{B-L} = 0$ due to the $B-L$
conservation of the Yukawa coupling, the logarithmic piece
automatically cancels, and the result agrees with the spurion method.

Using our result Eq.~(\ref{eq:m2ij}), we list in Table {\ref{table:LFV}}
the branching ratios of the $\mu \to e \gamma$ decay, the $\mu \to e$
conversion process in Al nuclei \cite{Kitano:2002mt}, and the $\tau \to
\mu \gamma$ decay using representative parameter sets point I and II
worked out in \cite{Kitano:2004zd} \footnote{Strictly speaking, the Fat
Higgs model \cite{Harnik:2003rs} requires an $O(1)$ mixing between the
composite and elementary Higgs fields, and the Yukawa couplings
$h_{i\alpha}$ need to be scaled by the mixing angle.}.
For other parameter sets, the branching ratios can be estimated by the
scaling of $m_{3/2}^{-4}$ and $\tan^2 \beta$.

\begin{table}
\begin{center}
\begin{tabular}{|l||l|l|} \hline\hline
& {\bf Point I}
& {\bf Point II} \\ 
& \ \ \ $\tan\beta = 0.9$
& \ \ \ $\tan\beta = 5$\\
& \ \ \ $m_{3/2} = 47$~TeV  
& \ \ \ $m_{3/2} = 142$~TeV  \\ \hline
$BR(\mu \to e \gamma)$ &
$1.6 \times 10^{-8} | h_{1\alpha} h_{2 \alpha}^* |^2 $&
$3.9 \times 10^{-7} | h_{1\alpha} h_{2 \alpha}^* |^2 $ \\ \hline
$BR(\mu \to e; {\rm Al} )$ &
$5.4 \times 10^{-10} | h_{1\alpha} h_{2 \alpha}^* |^2 $&
$8.9 \times 10^{-10} | h_{1\alpha} h_{2 \alpha}^* |^2 $ \\ \hline
$BR(\tau \to \mu \gamma)$ &
$8.3 \times 10^{-10} | h_{2\alpha} h_{3 \alpha}^* |^2 $&
$6.5 \times 10^{-8} | h_{2\alpha} h_{3 \alpha}^* |^2 $ \\ 
\hline\hline
\end{tabular}
\end{center}
\caption{ The branching ratios of the LFV processes are shown in the two
points of the parameter space chosen in \cite{Kitano:2004zd}. 
The branching ratios scale as $m_{3/2}^{-4}$ and $\tan^2 \beta$.
 } 
\label{table:LFV}
\end{table}

To get a sense on the size of LFV, we consider a simple neutrino mass
model based on flavor $U(1)$ symmetry that is consistent with
leptogenesis.  It assigns the flavor $U(1)$ charges $L_1(1)$, $L_2(0)$,
$L_3(0)$, $N_1(2)$, $N_2(1)$, $N_3(0)$.  The flavor $U(1)$ is assumed to
be broken by an order parameter $\epsilon \simeq 0.1$.  The Yukawa
matrix is
\begin{equation}
  \label{eq:Yukawa}
  h_{i\alpha} \simeq h_t \left( 
    \begin{array}{ccc}
      \epsilon^3 & \epsilon^2 & \epsilon\\
      \epsilon^2 & \epsilon & 1\\
      \epsilon^2 & \epsilon & 1
    \end{array} \right),
\end{equation}
while the right-handed neutrino masses are $M_1:M_2:M_3 \simeq
\epsilon^4:\epsilon^2:1$.  The top Yukawa coupling is approximately
$h_t (M_3) \simeq 0.6$ for $\tan\beta \gtrsim 5$.  The light neutrino
masses from the seesaw mechanism are
\begin{equation}
  \label{eq:mnu}
  m_\nu \propto \left( 
    \begin{array}{ccc}
      \epsilon^2 & \epsilon & \epsilon\\
      \epsilon & 1 & 1 \\
      \epsilon & 1 & 1
    \end{array} \right).
\end{equation}
This type of model was considered in \cite{Buchmuller:1996pa}, and can
successfully produce the observed baryon asymmetry from the decay of
$N_1$.  Note that the mass of $N_1$ is about $10^{10}$~GeV and is
allowed by the gravitino constraint Eq.~(\ref{eq:TRH}) from the
overclosure by the LSP.  We find $|h_{1\alpha} h_{2\alpha}^*| \simeq
h_t^2(M_3) \epsilon \simeq 0.036$ in this model
\cite{Sato:2000zh}.  Thus the branching ratios are roughly estimated
to be $10^{-11}$ (point I) and $10^{-10}$ (point II).
By taking into account the $O(1)$ ambiguity in the model parameters,
the predictions are comparable to the current experimental upper bound
$1.2 \times 10^{-11}$ \cite{Brooks:1999pu}.
%
%
%
%
%
%
In both points, observation of the $\mu \to e \gamma$ decay in the
planned experiments \cite{PSI, PRISM} is quite promising.
Also, the values for $\tau \rightarrow \mu \gamma$ and $\mu \rightarrow
e$ conversion are in the interesting range for on-going
\cite{Abe:2003sx, Brown:2002mp}, or future \cite{MECO, PRISM}
experiments, respectively.

The corresponding analyses have been done in the mSUGRA, and stringent
bounds on the model parameters are obtained since the logarithmic factor
in Eq.~(\ref{eq:mSUGRA}) gives $O(100)$ enhancement in $BR(\mu
\rightarrow e \gamma)$.
The situation is particularly severe in models with Yukawa unification
and flavor symmetries \cite{Sato:2000zh}. With fixing $\tan \beta$ and
the SU(2)$_L$ gaugino mass $M_2$ to be $\tan \beta = 3$ and $M_2 =
150$~GeV, the lower bound on the scalar electron mass is found to be
more than 1~TeV for the same Yukawa couplings as above.
%
The bound is more severe for larger values of $\tan \beta$.
Too large LFV is a generic feature in the mSUGRA with the seesaw model
when the right-handed neutrino scale is relatively high, unless we
assume a specific texture of the Yukawa matrix \cite{Ellis:2002fe} or a
cancellation among the diagrams \cite{Ellis:2001xt}.
It is interesting to recall that the universal scaler mass is introduced by
hand to avoid too large FCNC in the mSUGRA. However, the prescription is
insufficient in the seesaw model.
The situation is significantly improved in the anomaly mediation because
of the absence of the logarithmic factor.

In conclusion, we have presented a framework where SUSY flavor and CP
problems are automatically solved and the thermal leptogenesis is made
consistent with the gravitino constraint.  It relies on UV insensitive
anomaly-mediated supersymmetry breaking supplemented by the $D$-terms
for $U(1)_{B-L}$ and $U(1)_Y$.  The right-handed neutrino mass
explicitly breaks $U(1)_{B-L}$ and reintroduces the lepton flavor
violation (LFV), but it lacks the logarithmic enhancement unlike in
mSUGRA.  Therefore the size of LFV is easily consistent with the current
limits while it is already threatening in the mSUGRA for the same
neutrino parameters.
Also, calculation of the LFV processes is less ambiguous in the
framework, since the corrections to the slepton masses are independent
of physics above the mass scale of the right-handed neutrinos.

\begin{acknowledgments}
  HM and TY thank the organizers of the seesaw 04 workshop at KEK,
  especially Kenzo Nakamura, where the initial idea of the project
  emerged.  MI thanks the hospitality of the Institute for Advanced
  Study where the actual work started.  RK is the Marvin L.~Goldberger
  Member at the Institute for Advanced Study.  HM was supported
  by the Institute for Advanced Study, funds for Natural Sciences.
  This work was also supported in part by the DOE under contracts
  DE-FG02-90ER40542 and DE-AC03-76SF00098, and in part by NSF grant
  PHY-0098840.
\end{acknowledgments}


\end{document}